\documentclass[titlepage,12pt,a4paper,reqno]{article}
\usepackage{amsmath,amssymb,amsfonts,graphics,setspace,tensor}
\usepackage[linktocpage=true,colorlinks,pdftex]{hyperref}
\usepackage{xcolor}
\colorlet{linkequation}{blue}
\usepackage{bm}
\usepackage{doi}
\usepackage{hyperref}
\hypersetup{colorlinks, citecolor=violet, filecolor=black, linkcolor=black, urlcolor=blue}
\newcommand*{\refeq}[1]{%
  \begingroup
    \hypersetup{
      linkcolor=linkequation,
      linkbordercolor=linkequation,
    }%
    \ref{#1}%
  \endgroup
}
\allowdisplaybreaks

\begin{document}


\begin{titlepage}

\centerline{\Large \bf Gravitoelectromagnetic knot fields} 
\vskip 1cm
\centerline{\bf A. V. Cri\c{s}an$^{a}$,
C. R. L. Godinho$^{b}$ 
and I. V. Vancea$^{b}$
}
\vskip 0.5cm
\centerline{\sl $^{a}$Department of Mechanical Systems Engineering}
\centerline{\sl Technical University of Cluj-Napoca}
\centerline{\sl 103 – 105 Muncii Bld., C.P. 400641 Cluj-Napoca, Romania} 
\centerline{\sl $^{b}$Group of Theoretical Physics and Mathematical Physics,}
\centerline{\sl Department of Physics, Federal Rural University of Rio de Janeiro}
\centerline{\sl Cx. Postal 23851, BR 465 Km 7, 23890-000 Serop\'{e}dica - RJ,
Brazil}
\centerline{
\texttt{\small adina.crisan@mep.utcluj.ro;
ionvancea@ufrrj.br
} 
}
\vspace{0.5cm}

\centerline{27 February 2021}

\vskip 1.4cm
\centerline{\large\bf Abstract}
We construct a class of knot solutions of the gravitoelectromagnetic (GEM) equations in vacuum in the linearized gravity approximation by analogy with the Ra\~{n}ada-Hopf fields. For these solutions, the dual metric tensors of the bi-metric geometry of the gravitational vacuum with knot perturbations are given and the geodesic equation as a function of two complex parameters of the GEM knots are calculated. Finally, the Landau--Lifshitz pseudo-tensor and a scalar invariant of the GEM knots are computed.
\vskip 0.7cm 

\noindent

\end{titlepage}


\section{Introduction}

The gravitoelectromagnetism (GEM) formulation of gravity has proved to be a practical approach to studying gravitating systems and physical processes in the gravitational field. The GEM formalism based on the analogy at the classical level between gravity and electromagnetism is presented at its most efficient form in the linearized gravity  \cite{Mashhoon:2000jq,Kopeikin:2001dz} and Weyl tensor approaches \cite{Gemelli:2002wt,FilipeCosta:2006fz,Costa:2009nn,Costa:2012cw}. However, some more formal approaches to the GEM are also known, but that is currently under investigation \cite{Ramos:2010zza}.  

A very large range of fundamental problems have been addressed in the GEM framework. The generalization of the GEM concepts to the non-linear electromagnetism was studied in \cite{Duplij:2013qoa}, to the non-commutative geometry in \cite{Malekolkalami:2013fqa}, to higher dimensions in \cite{Kansu:2014oia}, to the teleparallel gravity in 
\cite{Spaniol:2014lba,Ming:2017tna} and to the non-local gravity in
\cite{Mashhoon:2019jkq}. By analogy with the quantization of the electromagnetic systems, several quantum aspects of the GEM were investigated in 
\cite{Kiefer:2004hv}-\cite{Santos:2019jxi}.
The gauge properties and the non-abelian extension of GEM formalism were discussed in \cite{Ramos:2020yvb,Clark:2000ff}.
The thermalization of the GEM systems and the research of several thermal quantum processes was performed in 
\cite{Santos:2016vfa}-\cite{Santos:2019svb}.
(For excellent reviews on the topic of the GEM formalism  and on its applications see, e.g., 
\cite{Jantzen:1992rg}-\cite{Poirier:2015cyu}).

This paper proposes a class of knot solutions of the GEM in vacuum equations, in the linearized gravity approach. These are new solutions parameterized by a pair of complex scalar maps $\phi$ and $\theta$ from $\mathbb{R} \times \mathbb{R}^{3}$ to $\mathbb{C}$ which describe the knot structure of the corresponding gravitomagnetic and gravitoelectric fields. The existence of GEM knot fields guaranteed by the formal equivalence between the GEM equations and the Maxwell equations for which knot solutions are known to exist for some time 
\cite{Trautman:1977im}-\cite{Ranada:1990}.
In the context of GEM, the knot fields bring novelty to the body of the known solutions of the linearly perturbed Einstein equations, in the sense that they describe a topologically non-trivial gravitational vacuum with the knot topology controlled by a set of complex scalar maps.

A general major motivation to study the topological abelian fields is to understand the relationship between their physical and mathematical properties. This is an active area of research in classical electromagnetism with significant progress made recently in this direction by the discovery and the analysis of a large set of new solutions of the field equations that generalize the electromagnetic Ra\~{n}ada fiels  in the linear 
\cite{Ranada:1997}-\cite{Kumar:2020xjr}
and non-linear electrodynamics 
\cite{Hoyos:2015bxa}-\cite{Nastase:2018gjs}.
From the mathematical point of view, several mathematical structures related to the topologically non-trivial fields have been explored in the recent literature among which are the twistors \cite{Dalhuisen:2012zz}, rational functions \cite{Lechtenfeld:2017tif,Kumar:2020xjr}, fibre bundles \cite{Arrayas:2017xvo}, space-time foliations \cite{Silva:2018ule} and generalized Finsler geometries \cite{Crisan:2020krc}. The interaction of electromagnetic knots with the matter for classical and quantum particles is discussed. 
\cite{Crisan:2020krc}-\cite{Crisan:2020isv} (for recent reviews on many of these results, see, e.g., 
\cite{Arrayas:2017sfq}-\cite{Crisan:2020a}). By generalizing the knot solutions to the gravitating system, one could largely explore the knowledge obtained so far in field theory to better understand the gravitational field's structure at least in the weak field approximation.

This work's primary goal is to give results of the generalization of the knot fields to gravity. Here, we focus on the linearized gravity away from sources in the GEM formulation, which originated in the works of Thirring and Lens \cite{Thirring:1918uj,Lense:1918zz}. The existence of the knot fields for gravitational and gravitating electromagnetic fields has been discussed in other research papers 
\cite{Swearngin:2013sks}-\cite{Cho:2013kra}.
In particular, the existence of the Hopf--Ra\~{n}ada solutions of the spin 2 field theory and their properties in the GEM formulation in terms of Weyl tensor were analyzed in  
\cite{Ramos:2018rdw}, \cite{Swearngin:2013sks}-\cite{Thompson:2014owa} and
\cite{Sabharwal:2019ngs}. In the GEM linearized gravity, the discussion of the Hopfions in the gravity/fluid correspondence was recently approached in \cite{Alves:2018wku}. 
For another interesting approach to the knot structure of vacuum see
\cite{Cho:2013kra}.

Since the GEM potentials are identified with some components of the linearized classical perturbation field in the harmonic gauge, the gravitational knots are interesting from mathematical and physical points of view.  The analogy between gravity and electromagnetism is used to construct the knotted field. However, the GEM fields' gravitational nature imposes typical constraints on the structure of these knots that do not have an electromagnetic equivalent. The existence of knots in the fluctuating gravitational vacuum induces a linearly perturbed space-time bi-metric geometrical structure. The two metrics are related to each other by the GEM equations' duality symmetry under the swap of the gravitoelectric and gravitomagnetic fields and represent a redundant description of the space-time geometry.  While this is a general property of the GEM induced geometry away from sources, in the knot fields, the relationship between the two metrics imposes a set of non-linear constraints on scalar complex maps that parametrize the gravitational knot field.  This paper constructs the GEM knot fields and starts the analysis of their geometrical and physical properties.

The paper is organized as follows. In Section 2, we give a concise review of the GEM formulation's fundamental relations in the linearized gravity following the reviews cited above, mainly \cite{Mashhoon:2000he, Mashhoon:2003ax,Bakopoulos:2014exa}. In Section 3, we construct the GEM knot fields in vacuum by analogy with the corresponding solutions to the Maxwell equations. A novelty here is the analysis of the GEM potentials in the polar representation of the complex scalars. Furthermore, we give the explicit form of the dual metric tensors of the perturbed space-time in terms of the scalar fields and calculate the geodesic equations. Since all the formulas are symmetric under the duality transformation of fields, we present our relation only for one of the scalars to avoid cluttering the formulas. Next, we calculate the Landau--Lifshitz pseudo-tensor for the GEM knots and compute a GEM knot field's scalar invariant. In the last section, we discuss our results and give some details on the GEM constraint for the static potentials. Furthermore, we present a shortlist of interesting topics on this subject.
Throughout this paper, the Minkowski metric with signature $\left( +, -, -, - \right)$ and the natural units $8 \pi G = c = 1$ are used.

\section{GEM formalism of linearized gravity}

In this section, we briefly review all the necessary relations of the GEM formalism following mainly\cite{Mashhoon:2000he, Mashhoon:2003ax,Bakopoulos:2014exa} and we establish our notations. 

The GEM is a formulation of the General Relativity in the weak field approximation in which an analogy between the Maxwell equations and the gauge fixed linearized Einstein equation is established. The main equations of the GEM are established by starting with small perturbations of the flat space-time metric defined in the standard way

\begin{equation}
g_{\mu \nu} \left(x \right ) = \eta {_{\mu \nu }}
+
h{_{\mu \nu }}
\left ( x \right )
\, ,
\label{metric-perturbation}
\end{equation}
where $\eta_{\mu\nu}$ is the Minkovski metric and  $\left| h_{\mu\nu} \right| \ll 1 $. In the decomposition given by Equation (\refeq{metric-perturbation}),  
the gravitational field is represented by a spin two tensor field $h_{\mu \nu}$ (classical graviton field) in the flat space-time background. The dynamics of the graviton field are obtained from the
linearized Einstein tensor. With the purpose to establish our notation, let us briefly review the relevant relations. The linearized Christoffel symbols, the Ricci tensor and the Ricci scalar have the following form \cite{Landau:1982dva}

\begin{align}
\Gamma{^{\alpha }}_{{\mu \nu }} 
& = 
\frac{1}{2}
\eta^{\alpha \rho }
\left(
\partial_{\nu} h_{\mu \rho} 
+ 
\partial_{\mu} h_{\nu \rho} 
-
\partial_{\rho} h_{\mu \nu}
\right)
\, , 
\label{Christoffel symbols}
\\
R_{\mu\nu} 
& =
\frac{1}{2}
\eta^{\rho \sigma}
\left( 
\partial_{\rho} \partial_{\nu} h_{ \sigma \mu}
+
\partial_{\rho} \partial_{\mu} h_{\sigma \nu}
-
\partial_{\rho} \partial_{\sigma} h{_{\mu\nu}}
- 
\partial_{\mu} \partial_{\nu} h_{\rho \sigma}
\right)
\label{Ricci tensor}
\, ,
\\
R & = 
\eta^{\mu \rho} \eta^{\nu \sigma}
\left(
\partial_{\nu} \partial_{\mu} h_{\rho \sigma} 
- 
\partial_{\mu} \partial_{\rho} h_{\nu \sigma}
\right)
\label{Ricci scalar}
\, .
\end{align}
The linearized Einstein tensor results from combining the equations above and takes the following standard form

\begin{equation}
G_{\mu\nu} 
= 
\frac{1}{2}
\left( 
\partial_{\alpha} \partial_{\nu} {h^{\alpha}}_{\mu} 
+
\partial_{\alpha} \partial_{\mu} {h^{\alpha}}_{\nu}
-
\partial^2 h_{\mu\nu}
- 
\partial_{\nu} \partial_{\mu} h
- 
\eta_{\mu\nu} \partial_{\beta} \partial_{\alpha} h^{\alpha\beta}
+
\eta_{\mu\nu} \, \partial^2 \, h
\right)
\label{Einstein Tensor}
\, ,
\end{equation}
where $\partial^{2} = \eta^{\mu \nu} \partial_{\mu} \partial_{\nu}$
and $h = \eta^{\mu \nu} h_{\mu \nu}$. The linearized Einstein equations are simplified by introducing the following trace-reversed metric perturbation 

\begin{equation}
\tilde{h}_{\mu\nu} =
h_{\mu\nu} - \frac{1}{2}
\eta_{\mu\nu}
h
\label{new perturbations}
\, .
\end{equation}
Then the linearized Einstein equations take the following form

\begin{equation}
\partial_{\alpha} \partial_{\nu} {\tilde{h}^{\alpha}}_{\: \: \: \mu}
+ 
\partial_{\alpha} \partial_{\mu} {\tilde{h}^{\alpha}}_{\: \: \: \nu}
- 
\partial^2 \tilde{h}_{\mu\nu}
- 
\eta_{\mu\nu} 
\partial_{\beta} \partial_{\alpha} {\tilde{h}^{\alpha\beta}}
= T_{\mu\nu}
\label{linearized--Einstein-trace-reversed}
\, ,
\end{equation}
where $T_{\mu \nu}$ is the matter energy-momentum tensor. As it is well known, the free graviton has a gauge symmetry given by the infinitesimal transformations

\begin{equation}
\delta_{\xi} h_{\mu} = \partial_{\mu} \xi_{\nu} + \partial_{\nu} \xi_{\mu}
\, ,
\label{gauge-symmetry}
\end{equation}
where $\xi$ is an infinitesimal arbitrary smooth vector field on flat space-time. By using this symmetry and fixing the gauge, the Einstein equations are simplified further. For example, in the trace-reversed harmonic gauge defined by the following equation

\begin{equation}
{\tilde{h}^{\mu \nu}}_{\: \: \: \: \: , \nu} = 0
\, ,
\end{equation}
the Einstein equations take the following simpler form
\begin{equation}
\partial^{2} \tilde{h}_{\mu \nu} = -2 T_{\mu \nu}
\, .
\label{linearized--Einstein}
\end{equation}

The analogy with the Maxwell equations which is at the core of the
GEM approach to the linearized gravity is made apparent by the following ansatz for the trace-reversed field $ \tilde{h}_{\mu\nu} $ \cite{Mashhoon:1993,Mashhoon:2000,Kopeikin:2001dz}

\begin{equation}
\tilde{h}_{0 \mu}= 4 \mathsf{A}_{\mu}
\, ,
\qquad
\tilde{h}_{ij}=0
\label{GEM-ansatze}
\, .
\end{equation}
The approximation above of the spatial components of $ \tilde{h}_{\mu\nu} $ is justified by the dependence of $\tilde{h}_{ij}$ on $c$ at order $O(c^{-4})$ while $\tilde{h}_{0\mu}$ runs as $O(c^{-2})$. 
\footnote{In the references \cite{FilipeCosta:2006fz,Bakopoulos:2014exa}, it was assumed that $\tilde{h}_{ij}$ take a different form from the one considered here, but this we will not discuss this hypothesis.}. The GEM potentials are written in the three dimensional form by identifying $\mathsf{A}^{\mu} = \left( \Phi , \textbf{A}/2 \right)$
\footnote{The numerical factors result from the identification of the GEM potentials with the corresponding components of the tensor  $ \tilde{h}_{\mu\nu} $ such that the equations obtained for $\mathsf{A}^{\mu}$ have the form of Maxwell's equations. For different numerical factors, the analogy between the electric permittivity of vacuum and $\varepsilon_G = 1/4 \pi G$ is lost.}. The harmonic gauge condition implies the following relations among the components of the GEM potential

\begin{align}
{\mathsf{A}^{\mu}}_{,\mu} = 0
\, ,
\qquad
{\tilde{h}^{ij}}_{\: \: \: \: , j} + {\mathsf{A}^{i}}_{, 0} = 0
\, .
\label{GEM-gauge-cond} 
\end{align}
The first equation above corresponds to the Lorentz gauge in the covariant form. The second equation is a typical GEM constraint and it has its origin in the gravitational structure that underlies the potentials. It implies that the spatial components of the potential $\mathsf{A}$ are independent of time for the ansatze given by Equations (\refeq{GEM-ansatze}) above. Due to this limitation it imposes on the type of solutions of the GEM equations and to the fact that it contains terms of order $O(c^{-4})$, the second constraint from (\refeq{GEM-gauge-cond}) is sometimes ignored \cite{Mashhoon:1993,Mashhoon:2000,Kopeikin:2001dz}. This point of view is adopted here and the properties of the general solutions are analyzed. However, the consequences of the GEM constraints are briefly discussed in the last section.

By analogy with the classical electromagnetism, the GEM electric and magnetic fields are introduced by the following combinations of derivatives of potentials

\begin{align}
\textbf{E}
& = 
- \nabla \Phi
-
\partial_{0}
\left(
\frac{\mathbf{A}}{2}
\right )
\label{GEM-E-field}
\\
\mathbf{B}
& =
\nabla \times
\left(
\frac{\mathbf{A}}{2}
 \right )
\, .
\label{GEM-B-field}
\end{align} 
It follows that the gravitomagnetic field and the gravitoelectric field
satisfy the Gauss and the Faraday--Lenz laws, respectively,

\begin{equation}
\partial_{i} E_{i} = 0
\, ,
\qquad
\varepsilon_{ijk} \partial_{j} E_{k} = - \partial_{0} B_{i}
\, .
\label{GEM-Gauss-B-Faraday}
\end{equation}
As is the case of the electromagnetism, the above equations are constraints on structure of the fields $\mathbf{E}$ and $\mathbf{B}$ that must be obeyed at all times. The dynamics
of $\mathbf{E}$ and $\mathbf{B}$ can be read off of the linearized Einsten equations (\refeq{linearized--Einstein}). In the three dimensional formulation, these equations have the following decomposition 

\begin{align}
\partial^{2} \mathsf{A}^{0} & = \partial^{2} \Phi  
= \frac{1}{2} T^{00}
\, ,
\label{GEM-eq-1}
\\
\partial^{2} \mathsf{A}^{i} & = \frac{1}{2} \partial^{2} A^{i}
= \frac{1}{2} T^{0i}
\, .
\label{GEM-eq-2}
\end{align}
Equations (\refeq{GEM-eq-1}) and (\refeq{GEM-eq-2}) are the Gauss law for the gravitoelectric field and the Maxwell--Amp\`{e}re law, respectively,

\begin{align}
\partial_{i} E_{i} 
& = \frac{1}{2} T_{00}
\, ,
\label{GEM-Gauss-E}
\\ 
\varepsilon_{ijk}\partial_{j} B_{k}
& =
\partial_{0}
E_{i}
+
\frac{1}{2} T_{0i}
\, .
\label{GEM-Ampere}
\end{align}
Using the electrodynamics tools to study the GEM fields' background, we gain valuable insight into various gravitating physical systems. The background has a pseudo-Riemann geometry, given by the line element of the form

\begin{equation}
ds^2 = 
\left(
1+2 \Phi
\right)
(d x^{0})^2 - 
4 \delta_{ij}
A^{i} dx^{0} d x^{j}
-
\left(
1 - 2 \Phi
\right)
\delta_{ij} dx^i dx^j
\, .
\label{space-time-line-element}
\end{equation}
The dynamics of the massive particles in the GEM background results from the classical geodesic equation

\begin{equation}
\frac{d^{2} x^{\mu}}{d s^{2}} 
+
{\Gamma^{\mu}}_{\nu \rho}
\frac{d x^{\nu}}{d s} 
\frac{d x^{\rho}}{d s}  
= 0
\, .
\label{GEM-geodesic}
\end{equation}
Note that the linearized Christoffel symbols calculated from the metric (\refeq{space-time-line-element}) should be employed in the equation in the (\refeq{GEM-geodesic}) above. For further details, we refer the reader to 
\cite{Mashhoon:2003ax}.

\section{General knot fields in linearized GEM}

IIn this section, are constructed a class of knot type solutions of the in-vacuum GEM equations and the space-time geometry generated by the GEM knots is also discussed.  These form a class of GEM fields parameterized by two scalar fields that satisfy some differentiability and regularity conditions, introduced by Ra\~{n}ada in the context of the electromagnetism in \cite{Ranada:1989wc,Ranada:1990}. A different set of knotted GEM fields was obtained in the GEM formulation in terms of Weyl tensor in \cite{Swearngin:2013sks,Thompson:2014,Thompson:2014owa}.

\subsection{Knot fields in linearized GEM}

In order to explore the knot like solutions in the GEM formalism in the linearized gravity reviewed in the previous section, we need to consider the GEM field away from its sources for which the field equations are (\refeq{GEM-Gauss-B-Faraday}), (\refeq{GEM-Gauss-E}) and (\refeq{GEM-Ampere}) in vacuum. In this case, the knot-like solutions are confirmed by the analogy with the corresponding Maxwell equations that admit topologically non-trivial solutions with knot structure, which were given for the first time in \cite{Trautman:1977im,Ranada:1989wc,Ranada:1990}.  

In the absence of the matter fields, the set of GEM equations (\refeq{GEM-Gauss-B-Faraday}), (\refeq{GEM-Gauss-E}) and (\refeq{GEM-Ampere}) is invariant under the gravitoelectromagnetic duality transformation $ E_{i} \to B_{i}$ and $B_{i} \to - E_{i}$. That allows one to describe the fields $\mathbf{E}$ and $\mathbf{B}$ symmetrically in terms of two potentials $\mathsf{A}$ and $\mathsf{C}$ which help visualize the line structure of each field. Since this is a redundant description of the GEM field, there is a relationship between $\mathsf{A}$ and $\mathsf{C}$. Let us construct the general knot solutions of the GEM equations.

Let us start by taking two smooth complex scalar maps $\phi , \theta : \mathbb{R} \times \mathbb{R}^{3} \to \mathbb{C}$ whose level equations $\phi (x) = \phi_{0} = \text{constant}$ and $\theta (x) = \theta_{0} = \text{constant}$ are associated to gravitomagnetic and gravitoelectric field lines\footnote{The Hopf knots are given by maps from $S^{3}$ to $S^{2}$. We will comment on the particularities of this case in the last section.}. The functions $\phi$ and $\theta$ are assumed to be regular at spatial infinity and they define the following pair of potentials \cite{Vancea:2017tmx}

\begin{align}
\mathsf{A}_{\mu} & = \frac{\sqrt{a}}{4 \pi i}
\left(
\frac{\bar{\phi} \partial_{\mu} \phi - \phi \partial_{\mu} \bar{\phi}}{1+ \vert \phi \vert^{2}}
\right)
\, ,
\label{GEM-A-potentials}
\\
\mathsf{C}_{\mu} & = \frac{\sqrt{a}}{4 \pi i}
\left(
\frac{\bar{\theta} \partial_{\mu} \theta - \theta \partial_{\mu} \bar{\theta}}{1+ \vert \theta \vert^{2}}
\right)
\, ,
\label{GEM-C-potentials}
\end{align}
where $a$ is a dimensionful constant. Since the two potentials must describe the same fields $\mathbf{E}$ and $\mathbf{B}$, there is the relationship between $\mathsf{A}$ and $\mathsf{C}$ that has the following form in terms of scalar fields

\begin{align}
\left(
1 + \vert \phi \vert^{2}
\right)^{-2}
\varepsilon_{ijk} \partial_{j} \phi \partial_{k} \bar{\phi}
& = 
\left(
1 + \vert \theta \vert^{2}
\right)^{-2}
\left(
\partial_{0} \bar{\theta} \partial_{i} \theta
-
\partial_{0} \theta \partial_{i} \bar{\theta}
\right)
\, ,
\label{GEM-relation-phi-theta-1}
\\
\left(
1 + \vert \theta \vert^{2}
\right)^{-2}
\varepsilon_{ijk} \partial_{j} \theta \partial_{k} \bar{\theta}
& = 
\left(
1 + \vert \phi \vert^{2}
\right)^{-2}
\left(
\partial_{0} \bar{\phi} \partial_{i} \phi
-
\partial_{0} \phi \partial_{i} \bar{\phi}
\right)
\, .
\label{GEM-relation-phi-theta-2}
\end{align}
The potentials defined by Equations (\refeq{GEM-A-potentials})
and (\refeq{GEM-C-potentials}) generate the following knotted fields
that is in vacuum solutions to the GEM equations 

\begin{align}
E_{i} & = \frac{\sqrt{a}}{2 \pi i } \:
\frac{
\varepsilon_{ijk} \partial_{j} \theta \partial_{k} \bar{\theta}
}
{\left( 1 + \vert \theta \vert^{2} \right)^{2}}
\, ,
\label{GEM-knot-electric}
\\
B_{i} & = \frac{\sqrt{a}}{2 \pi i } \:
\frac{
\varepsilon_{ijk} \partial_{j} \phi \partial_{k} \bar{\phi}
}
{\left( 1 + \vert \phi \vert^{2} \right)^{2}}
\, .
\label{GEM-knot-magnetic}
\end{align}
The relations (\refeq{GEM-relation-phi-theta-1}) and (\refeq{GEM-relation-phi-theta-2}) show that a single potential is sufficient to describe the gravitomagnetic and the gravitoelectric field. For example, by employing only $\phi$, we have the following representation of $\mathbf{E}$ and $\mathbf{B}$

\begin{align}
E_{i} & = \frac{\sqrt{a}}{2 \pi i} \:
\frac{\partial_{0} \bar{\phi} \partial_{i} \phi 
- 
\partial_{0} \phi \partial_{i} \bar{\phi}}
{\left( 1 + \vert \phi \vert^{2} \right)^{2}}
\, ,
\label{GEM-knot-electric-phi}
\\
B_{i} & = \frac{\sqrt{a}}{2 \pi i } \:
\frac{
\varepsilon_{ijk} \partial_{j} \phi \partial_{k} \bar{\phi}
}
{\left( 1 + \vert \phi \vert^{2} \right)^{2}}
\, .
\label{GEM-knot-magnetic-phi}
\end{align}
The two potentials help to visualize the field lines of each field 
and display the duality symmetry of the GEM equations. The above solutions form a class of knot type GEM fields because they are defined in rather general complex functions. By taking $\phi$ and $\theta$ of particular forms, sets of solutions with specific features are constructed (see for a recent review \cite{Arrayas:2017sfq}). 

In order to analyze the properties of the GEM knot fields and the induced space-time geometry, it is convenient to pass to the following polar representation of the scalar fields

\begin{align}
\phi (x) & =  \rho (x) e^{i \alpha (x)}
\, ,
\qquad
\bar{\phi} (x) =  \rho (x) e^{-i \alpha (x)}
\, ,
\label{GEM-polar-phi}
\\
\theta (x) & =  \lambda (x) e^{i \beta (x)}
\, ,
\qquad
\bar{\theta} (x) =  \lambda (x) e^{-i \beta (x)}
\, ,
\label{GEM-polar-theta}
\end{align}
where $\rho$ and $\lambda$ are positively defined functions everywhere and obey regularity conditions at infinity. In this representation, the GEM potentials take the following form

\begin{align}
\mathsf{A}_{\mu} & = \frac{\sqrt{a}}{2 \pi} \frac{\rho^{2} \partial_{\mu} \alpha}{\left( 1+ \rho^{2}\right)}
\, ,
\label{GEM-polar-A}
\\
\mathsf{C}_{\mu} & = \frac{\sqrt{a}}{2 \pi} \frac{\lambda^{2} \partial_{\mu} \beta}{\left(1+ \lambda^{2}\right)}
\, .
\label{GEM-polar-C}
\end{align}
From the relations (\refeq{GEM-polar-A}) and (\refeq{GEM-polar-C}) we derive the gravitoelectric and gravitomagnetic fields in the following form

\begin{align}
E_{i} 
& = 
- \frac{\sqrt{a}}{\pi}
\left[
\frac{\rho 
\left(
\partial_{0} \rho \partial_{i} \alpha
+
\partial_{i} \rho \partial_{0} \alpha
\right)
}{
\left(
1 + \rho^{2}
\right)^{2}
}
+
\frac{\rho^{2} 
\partial_{0} \rho \partial_{i} \alpha
}{
1 + \rho^{2}
}
\right]
\, ,
\label{GEM-polar-E}
\\
B_{i} 
& = 
\frac{\sqrt{a}}{\pi}
\varepsilon_{ijk}
\frac{\rho 
\partial_{j} \rho \partial_{k} \alpha
}{
\left(
1 + \rho^{2}
\right)^{2}
}
\, .
\label{GEM-polar-B}
\end{align}
The relations (\refeq{GEM-relation-phi-theta-1}) and (\refeq{GEM-relation-phi-theta-2}) between the complex fields take the following form in polar representation

\begin{align}
\frac{\rho}{\left(1 + \rho^{2} \right)^{2}}
\varepsilon_{ijk}
\left(
\partial_{j} \rho \partial_{k} \alpha
-
\partial_{j} \alpha \partial_{k} \rho
\right)
& =
\frac{2 \lambda}{\left( 1 + \lambda^{2} \right)^{2}}
\left(
\partial_{i} \lambda \partial_{0} \beta
-
\partial_{0} \lambda \partial_{i} \beta
\right)
\, ,
\label{GEM-relation-phi-polar-1}
\\
\frac{\lambda}{\left(1 + \lambda^{2} \right)^{2}}
\varepsilon_{ijk}
\left(
\partial_{j} \lambda \partial_{k} \beta
-
\partial_{j} \beta \partial_{k} \lambda
\right)
& =
\frac{2 \rho}{\left( 1 + \rho^{2} \right)^{2}}
\left(
\partial_{i} \rho \partial_{0} \alpha
-
\partial_{0} \rho \partial_{i} \alpha
\right)
\, .
\label{GEM-relation-phi-polar-2}
\end{align}
Next, a few comments are required.  As mentioned above, the fact that the fields defined by Equations (\refeq{GEM-knot-electric}) and (\refeq{GEM-knot-magnetic}) are solutions of the in-vacuum GEM equations is a consequence of the formal equivalence between the Maxwell and the GEM equations, respectively.  However, there is an important particularity of the gravitoelectric and gravitomagnetic fields which is a result of their gravitational origin, namely, the constraints on $\mathbf{E}$ and $\mathbf{B}$ given by the equations from (\refeq{GEM-gauge-cond}) above. Of particular interest here is the second GEM constraint on which we will comment in the last section. Furthermore, it is known that in the electromagnetism, new solutions are generated from a given knot field by using the invariance of the Maxwell equations under Lorentz and conformal symmetries or by working on specific space-time manifolds. These methods might not be applicable to GEM knots due to the underlying graviton structure \cite{Arrayas:2017sfq,Lechtenfeld:2017tif,Kumar:2020xjr}.

\subsection{Space-time geometry from GEM knots}

Interesting information about the space-time geometry generated by GEM knots away from the matter distribution is obtained by calculating the relevant tensorial objects. Since the geometry is of pseudo-Riemann type, its fundamental tensors are obtained from the metric tensor given by the line element from Equation (\refeq{space-time-line-element}) above. 
 
One important characteristic of the knot solutions is that they generate a bi-metric geometry on space-time with the two metrics depending on each other according to the relations (\refeq{GEM-relation-phi-theta-1}) and (\refeq{GEM-relation-phi-theta-2}). As we have seen before, this is a consequence of the gravitoelectromagnetic duality of the vacuum equations. The bi-metric geometry has one metric generated by the field associated with the potential $\mathsf{A}$ and the second (dual) metric associated with the potential $\mathsf{C}$. The dual metric is obtained by replacing $\mathsf{A}$ by $\mathsf{C}$ in Equation (\refeq{space-time-line-element}). Explicitly, we have the following metric tensors

\begin{align}
g_{\mu \nu} & =
\begin{bmatrix} 
1 +  \frac{\sqrt{a}}{\pi} \frac{\rho^{2}\partial_{0} \alpha}{1 + \rho^{2}}
& 
- \frac{4 \sqrt{a}}{\pi} \frac{\rho^{2}\partial_{1} \alpha}{1 + \rho^{2}}
& 
- \frac{4 \sqrt{a}}{\pi} \frac{\rho^{2}\partial_{2} \alpha}{1 + \rho^{2}} 
& 
- \frac{4 \sqrt{a}}{\pi} \frac{\rho^{2}\partial_{3} \alpha}{1 + \rho^{2}}
\\
- \frac{4 \sqrt{a}}{\pi} \frac{\rho^{2}\partial_{1} \alpha}{1 + \rho^{2}} 
&
- 1 +  \frac{\sqrt{a}}{\pi} \frac{\rho^{2}\partial_{0} \alpha}{1 + \rho^{2}}
&
0 
& 
0
\\
- \frac{4 \sqrt{a}}{\pi} \frac{\rho^{2}\partial_{2} \alpha}{1 + \rho^{2}} 
&
0
&
- 1 +  \frac{\sqrt{a}}{\pi} \frac{\rho^{2}\partial_{0} \alpha}{1 + \rho^{2}}
&
0
\\
- \frac{4 \sqrt{a}}{\pi} \frac{\rho^{2}\partial_{3} \alpha}{1 + \rho^{2}}
&
0
&
0
&
- 1 +  \frac{\sqrt{a}}{\pi} \frac{\rho^{2}\partial_{0} \alpha}{1 + \rho^{2}}
\end{bmatrix}
\, ,
\label{GEM-g-metric-tensors}
\\
f_{\mu \nu} & =
\begin{bmatrix} 
1 +  \frac{\sqrt{a}}{\pi} \frac{\lambda^{2}\partial_{0} \beta}{1 + \lambda^{2}}
& 
- \frac{4 \sqrt{a}}{\pi} \frac{\lambda^{2}\partial_{1} \beta}{1 + \lambda^{2}}
& 
- \frac{4 \sqrt{a}}{\pi} \frac{\lambda^{2}\partial_{2} \beta}{1 + \lambda^{2}} 
& 
- \frac{4 \sqrt{a}}{\pi} \frac{\lambda^{2}\partial_{3} \beta}{1 + \lambda^{2}}
\\
- \frac{4 \sqrt{a}}{\pi} \frac{\lambda^{2}\partial_{1} \beta}{1 + \lambda^{2}} 
&
- 1 +  \frac{\sqrt{a}}{\pi} \frac{\lambda^{2}\partial_{0} \beta}{1 + \lambda^{2}}
&
0 
& 
0
\\
- \frac{4 \sqrt{a}}{\pi} \frac{\lambda^{2}\partial_{2} \beta}{1 + \lambda^{2}} 
&
0
&
- 1 +  \frac{\sqrt{a}}{\pi} \frac{\lambda^{2}\partial_{0} \beta}{1 + \lambda^{2}}
&
0
\\
- \frac{4 \sqrt{a}}{\pi} \frac{\lambda^{2}\partial_{3} \beta}{1 + \lambda^{2}}
&
0
&
0
&
- 1 +  \frac{\sqrt{a}}{\pi} \frac{\lambda^{2}\partial_{0} \beta}{1 + \lambda^{2}}
\end{bmatrix}
\, .
\label{GEM-f-metric-tensors}
\end{align}
The relationship between the two dual geometries is an open problem. The main difficulty in determining the explicit relations between geometrical objects, e.g., the two metric tensors, resides in the non-linearity of Equations (\refeq{GEM-relation-phi-theta-1}) and (\refeq{GEM-relation-phi-theta-2}) that makes it hard from the analytical point of view to define a duality map between the two geometries. Indeed, as Equations (\refeq{GEM-g-metric-tensors}) and (\refeq{GEM-f-metric-tensors}) show, the scalar functions enter non-linearly in the metric tensors $g_{\mu \nu}$ and $f_{\mu \nu}$. The main difficulty is to solve Equations (\refeq{GEM-relation-phi-theta-1}) and (\refeq{GEM-relation-phi-theta-2}) and to find out an explicit relationship between $g_{\mu \nu}$ and $f_{\mu \nu}$. The same problem occurs in the Ra\~{n}ada construction. 

Let us focus on the geodesic equation in the geometry determined by the metric tensor $g_{\mu \nu}$. The dual geometry calculations follow precisely the same line, and avoid cluttering formulas will not be displayed here. In determining the geodesic equation in the GEM knot background, are needed the Christoffel symbols. They result from Equations (\refeq{Christoffel symbols}) and (\refeq{GEM-polar-A}) above. The following expressions result after applying some straightforward algebra

\begin{align}
{\Gamma^{0}}_{00}
& =
\frac{\sqrt{a}}{2 \pi}
\left[
\frac{2 \rho 
\partial_{0} \rho \partial_{0} \alpha
}{
\left(
1 + \rho^{2}
\right)^{2}
}
+
\frac{\rho^{2} 
\partial^{2}_{0} \alpha
}{
1 + \rho^{2}
}
\right]
\, ,
\label{GEM-Christoffel-1} 
\\
{\Gamma^{0}}_{0i}
& =
\frac{\sqrt{a}}{2 \pi}
\left[
\frac{2 \rho 
\partial_{0} \rho \partial_{i} \alpha
}{
\left(
1 + \rho^{2}
\right)^{2}
}
+
\frac{\rho^{2} 
\partial_{0} \partial_{i} \alpha
}{
1 + \rho^{2}
}
\right]
\, ,
\label{GEM-Christoffel-2} 
\\
{\Gamma^{0}}_{ij}
& =
\frac{\sqrt{a}}{\pi}
\left[
\frac{\rho 
\left(
\partial_{i} \rho \partial_{j} \alpha
+
\partial_{j} \rho \partial_{i} \alpha
\right)
}{
\left(
1 + \rho^{2}
\right)^{2}
}
+
\frac{\rho^{2}
\partial_{i} \partial_{j} \alpha
}{
1 + \rho^{2}
}
\right]
\, ,
\label{GEM-Christoffel-3}
\\
{\Gamma^{i}}_{00}
& =
\frac{\sqrt{a}}{2 \pi}
\delta^{ik}
\left[
\frac{
2 \rho 
\left(
\partial_{k} \rho \partial_{0} \alpha
-
4 
\partial_{0} \rho \partial_{k} \alpha
\right)
}{
\left(
1 + \rho^{2}
\right)^{2}
}
-
\frac{3 \rho^{2} 
\partial_{k} \partial_{0} \alpha
}{
1 + \rho^{2}
}
\right]
\, ,
\label{GEM-Christoffel-4} 
\\
{\Gamma^{i}}_{0j}
& =
- 
\frac{\sqrt{a}}{\pi}
\delta^{ik}
\left[
\frac{
2 \rho 
\left(
\partial_{k} \rho \partial_{j} \alpha
-
\partial_{j} \rho \partial_{k} \alpha
\right)
}{
\left(
1 + \rho^{2}
\right)^{2}
}
\right]
-
\frac{\sqrt{a}}{2 \pi}
{\delta^{i}}_{j}
\left[
\frac{2 \rho 
\partial_{0} \rho \partial_{0} \alpha
}{
\left(
1 + \rho^{2}
\right)^{2}
}
+
\frac{\rho^{2} 
\partial^{2}_{0} \alpha
}{
1 + \rho^{2}
}
\right]
\, ,
\label{GEM-Christoffel-5}
\\
{\Gamma^{i}}_{jk}
& =
\frac{\sqrt{a}}{2 \pi}
\delta^{im}
\left[
\frac{
2 \rho 
\left(
\delta_{jk}
\partial_{m} \rho \partial_{0} \alpha
-
\delta_{jm}
\partial_{k} \rho \partial_{0} \alpha
\right)
}{
\left(
1 + \rho^{2}
\right)^{2}
}
+
\frac{\rho^{2} 
\left(
\delta_{jk}
\partial_{m} \partial_{0} \alpha
-
\delta_{jm}
\partial_{k} \partial_{0} \alpha
\right)
}{
1 + \rho^{2}
}
\right]
\nonumber
\\
& -
\frac{\sqrt{a}}{2 \pi}
\delta^{im}
\delta_{km}
\left[
\frac{2 \rho 
\partial_{j} \rho \partial_{0} \alpha
}{
\left(
1 + \rho^{2}
\right)^{2}
}
+
\frac{\rho^{2} 
\partial_{j} \partial_{0} \alpha
}{
1 + \rho^{2}
}
\right]
\, .
\label{GEM-Christoffel-6}
\end{align} 
The geodesic equation can be easily obtained from the above equations. For completeness, we give it here in the component form
\begin{align}
\ddot{x}^{0}
&
+
\frac{\sqrt{a}}{2\pi}
\left[
\frac{2 \rho 
\partial_{0} \rho \partial_{0} \alpha
}{
\left(
1 + \rho^{2}
\right)^{2}
}
+
\frac{\rho^{2} 
\partial_{0} \partial_{0} \alpha
}{
1 + \rho^{2}
}
\right]
\cdot
\dot{x}^{0} \dot{x}^{0}
+
\frac{\sqrt{a}}{\pi}
\left[
\frac{2 \rho 
\partial_{0} \rho \partial_{i} \alpha
}{
\left(
1 + \rho^{2}
\right)^{2}
}
+
\frac{\rho^{2} 
\partial_{0} \partial_{i} \alpha
}{
1 + \rho^{2}
}
\right]
\dot{x}^{0} \dot{x}^{i}
\nonumber
\\
&
+
\frac{\sqrt{a}}{2 \pi}
\left[
\frac{2 \rho 
\left(
\partial_{i} \rho \partial_{j} \alpha
+
\partial_{j} \rho \partial_{i} \alpha
\right)
}{
\left(
1 + \rho^{2}
\right)^{2}
}
+
\frac{\rho^{2} 
\left(
\partial_{i} \partial_{j} \alpha
+
\partial_{j} \partial_{i} \alpha
\right)
}{
1 + \rho^{2}
}
\right]
\dot{x}^{i} \dot{x}^{j}
=
0
\, ,
\label{GEM-geodesic-equation-1}
\\
\ddot{x}^{i}
&
+
\frac{\sqrt{a}}{2 \pi}
\delta^{ik}
\left[
\frac{
2 \rho 
\left(
\partial_{k} \rho \partial_{0} \alpha
-
4 
\partial_{0} \rho \partial_{k} \alpha
\right)
}{
\left(
1 + \rho^{2}
\right)^{2}
}
+
\frac{\rho^{2} 
\left(
\partial_{k} \partial_{0} \alpha
-
4
\partial_{0} \partial_{k} \alpha
\right)
}{
1 + \rho^{2}
}
\right]
\:
\dot{x}^{0} \dot{x}^{0}
\nonumber
\\
&
-
2
\frac{\sqrt{a}}{\pi}
\delta^{ik}
\left[
\frac{
2 \rho 
\left(
\partial_{k} \rho \partial_{j} \alpha
-
\partial_{j} \rho \partial_{k} \alpha
\right)
}{
\left(
1 + \rho^{2}
\right)^{2}
}
+
\frac{\rho^{2} 
\left(
\partial_{k} \partial_{j} \alpha
-
\partial_{j} \partial_{k} \alpha
\right)
}{
1 + \rho^{2}
}
\right]
\:
\dot{x}^{0} \dot{x}^{j}
\nonumber
\\
& -
\frac{\sqrt{a}}{\pi}
{\delta^{i}}_{j}
\left[
\frac{2 \rho 
\partial_{0} \rho \partial_{0} \alpha
}{
\left(
1 + \rho^{2}
\right)^{2}
}
+
\frac{\rho^{2} 
\partial_{0} \partial_{0} \alpha
}{
1 + \rho^{2}
}
\right]
\:
\dot{x}^{0} \dot{x}^{j}
\nonumber
\\
&
+
\frac{\sqrt{a}}{2 \pi}
\delta^{im}
\left[
\frac{
2 \rho 
\left(
\delta_{jk}
\partial_{m} \rho \partial_{0} \alpha
-
\delta_{jm}
\partial_{k} \rho \partial_{0} \alpha
\right)
}{
\left(
1 + \rho^{2}
\right)^{2}
}
+
\frac{\rho^{2} 
\left(
\delta_{jk}
\partial_{m} \partial_{0} \alpha
-
\delta_{jm}
\partial_{k} \partial_{0} \alpha
\right)
}{
1 + \rho^{2}
}
\right]
\:
\dot{x}^{j} \dot{x}^{k}
\nonumber
\\
& -
\frac{\sqrt{a}}{2 \pi}
\delta^{im}
\delta_{km}
\left[
\frac{2 \rho 
\partial_{j} \rho \partial_{0} \alpha
}{
\left(
1 + \rho^{2}
\right)^{2}
}
+
\frac{\rho^{2} 
\partial_{j} \partial_{0} \alpha
}{
1 + \rho^{2}
}
\right]
\:
\dot{x}^{j} \dot{x}^{k}
=
0
\, .
\label{GEM-geodesic-equation-2}
\end{align}
Equations (\refeq{GEM-geodesic-equation-1}) and (\refeq{GEM-geodesic-equation-2}) describe the dynamics of a test particle of given mass in the GEM knot background.

\subsection{Energy-momentum pseudo-tensor and scalar invariant}

The physical and geometrical properties of the GEM knot fields are described in terms of several tensor and scalar objects. One important quantity that contains information similar to the Maxwell tensor and the Poynting vector of the electromagnetic field is the Landau--Lifshitz pseudo-tensor. The general form of it in the GEM formalism was given in \cite{Mashhoon:2000he}. Here, we will discuss its form for static GEM knots with $\partial_{0} \mathsf{A}_{\mu} = 0$. After some algebra, we obtain the following relations for the Landau--Lifshitz tensor

\begin{align}
t_{00} 
& =
- \frac{a}{(4 \pi)^{2}
\left(
1 + \rho^{2}
\right)^{4}
}
\left[
\frac{4 \rho^{2}}{
\left(
1 + \rho^{2}
\right)^{4}}
\left(
\partial_{(i} \rho \partial_{j)} \alpha
\right)
\left(
\partial^{(i} \rho \partial^{j)} \alpha
\right)
+
\frac{8 \rho^{3}}
{\left(
1 + \rho^{2}
\right)^{3}}
\partial_{(i} \rho \partial_{j)} \alpha
\,
\partial^{i} \partial^{j} \alpha
\right.
\nonumber
\\
&+
\left.
\frac{4 \rho^{2}}
{\left(
1 + \rho^{2}
\right)^{2}}
\partial_{i} \partial_{j} \alpha
\,
\partial^{i} \partial^{j} \alpha
\right]
\, , 
\label{GEM-Landau-Lifshitz-tensor-1}
\\
t_{0i} & = 0
\, ,
\label{GEM-Landau-Lifshitz-tensor-2}
\\
t_{ij} & =
\frac{a \rho^{4}}{\pi^{2} \left( 1 + \rho^{2} \right)^{6}} 
\left[
\varepsilon_{imn} \varepsilon_{jrs} 
\partial^{m} \rho \partial^{r} \rho 
\partial^{n} \alpha \partial^{s} \alpha
+
\delta_{ij}
\left(
\partial^{m} \rho \partial_{m} \rho
\right)
\left(
\partial^{n} \alpha \partial_{n} \alpha
\right)
-
\left( 
\partial^{m} \rho \partial_{m} \alpha
\right)^{2}
\right]
\label{GEM-Landau-Lifshitz-tensor-3}
\, .
\end{align}
From the above equations, we see that the GEM Poynting vector $S_{i} =  \epsilon_{ijk} E_{i} B_{k}$ vanishes since for static GEM knots $t_{0i} = 4 \epsilon_{ijk} E_{i} B_{k}$. Thus, only the gravitomagnetic field contributes to the spatial components of the Landau--Lifshitz pseudo-tensor. However, in the dual picture, only the gravitoelectric field enters the Maxwell tensor analog while the GEM Poynting vectors continue to be zero.

The analogy between linearized gravity and electromagnetism provides more information about the scalar invariants' GEM fields. Let us calculate the invariant $F^{2}$ that depends on the first derivative of the $ \tilde{h}_{\mu\nu} $. This invariant was discussed for more general configurations in \cite{Bakopoulos:2016rkl}. The general definition of $F^{2}$ is
\begin{equation}
F^{2} = F^{\alpha \mu \nu} F_{\alpha \mu \nu}
\, ,
\label{GEM-F-definition}
\end{equation} 
where $F_{\mu \nu \rho} $ is obtained from the linearized Einstein tensor and has the following form

\begin{equation}
F_{\alpha \mu \nu} 
= 
\partial_{\nu} \tilde{h}_{\alpha \mu} 
+
\partial_{\mu} \tilde{h}_{\alpha \nu}
-		
\partial_{\alpha} \tilde{h}_{\mu \nu}
-
\eta_{\mu \nu} \partial^{\beta} \tilde{h}_{\alpha \beta}
\, .
\label{GEM-F-tensor}
\end{equation}
For the GEM knots discussed here, the scalar $F^{2}$ takes the following form

\begin{align}
\frac{1}{4}
F^{2}
& = 
- \frac{16 a \rho^{2}}{\pi^{2} \left( 1 + \rho^{2} \right)^{4}}
\left[
\left(
\partial^{m} \rho \partial_{m} \rho
\right)
\left(
\partial^{n} \alpha \partial_{n} \alpha
\right)
-
\left( 
\partial^{m} \rho \partial_{m} \alpha
\right)^{2}
\right]
\nonumber
\\
& -
\frac{8 a}{\pi^{2}}
\left[
\frac{\rho 
\left(
\partial_{0} \rho \partial_{m} \alpha
+
\partial_{i} \rho \partial_{0} \alpha
\right)
}{
\left(
1 + \rho^{2}
\right)^{2}
}
+
\frac{\rho^{2} 
\partial_{0} \rho \partial_{m} \alpha
}{
1 + \rho^{2}
}
\right]^{2}
\nonumber
\\
& -
\frac{a}{4 \pi^{2} \left(
 1+\rho^{2}
 \right)^{4}}
 \left\{
 \left[
\frac{2\rho\partial_{m} \rho \partial_{0} \alpha}
{\left(1+ \rho^{2}
 \right )^{2}}
+\frac{\rho^{2} \partial_{m} \partial_{0} \alpha}{1+\rho^{2}}
 \right ]^{2}
+
\left[
\frac{2\rho\partial^{m} \rho \partial_{m} \alpha}
{\left(1+ \rho^{2}
 \right )^{2}}
+\frac{\rho^{2} \partial^{m} \partial_{m} \alpha}{1+\rho^{2}}
 \right ]^{2} 
\right\}
\nonumber
\\
& +
\frac{a}{4 \pi^{2} \left(
 1+\rho^{2}
 \right)^{4}}
\left[
\frac{2\rho
\left(
\partial_{m} \rho \partial_{n} \alpha
+
\partial_{n} \rho \partial_{m} \alpha
\right)
}
{\left(1+ \rho^{2}
\right )^{2}}
+
\frac{\rho^{2} 
\left(
\partial_{m} \partial_{n} \alpha
+
\partial_{n} \partial_{m} \alpha
\right)
}{1+\rho^{2}}
 \right ]^{2}
 \, ,
\label{GEM-F-scalar-knot}
\end{align}
where $v^{2} = v^{m} v_{m} = \delta^{mn} v_{m} v_{n}$ is the spatial norm.

The results obtained in this section show that the general GEM knot fields' relevant quantities are calculable, but the formulas obtained are in general non-linear and non-polynomial in terms of field line maps. Therefore, in concrete applications, computer-assisted methods and approximations are necessary to extract the physical information from these relations. A complete analysis of the energy-momentum tensor along the line of \cite{Mashhoon:1998tt} could be made here, but this is out of the present study's scope.

\section{Discussions}

This paper has investigated a class of knot solutions of the in-vacuum GEM equations  and discuss some geometrical and physical properties. We have shown that the knot fields induce a bi-metric geometry of the perturbed space-time with the two metric tensors $g_{\mu \nu}$ and $f_{\mu \nu}$ related to each other in a complicated way. The derivation of an explicit relation between the two metric tensors is challenging due to the non-linear equations that connect the two scalar functions that determine these tensors according to Equations (\refeq{GEM-g-metric-tensors}) and (\refeq{GEM-f-metric-tensors}). This is just a dual description of the space-time geometry due to the electric-magnetic duality of the GEM equations in the absence of matter. 
In one of these descriptions, the field lines of the gravitoelectric component of the linearized gravitational tensor $\tilde{h}_{\mu \nu}$ are given by the level lines of the scalar field $\theta$, while in the dual geometry the field lines of the gravitomagnetic components are identified with the level lines of the field $\phi$. All this suggests that away from matter sources, one describes the GEM field in two complementary ways. Nevertheless, since only one GEM potential is necessary to solve the motion equations, the two descriptions based on the two potentials $A$ and $C$ should be physically equivalent.

Another result obtained here is the derivation of the geodesic equation that describes a test particle's motion in a GEM knot field in one of the geometries and calculates the Landau--Lifschitz tensor and a scalar invariant of the GEM knots.

A critical feature of the GEM knot fields obtained here is that they are parametrized by two complex scalar maps $\phi$ and $\theta$. However, an essential subtlety arises when the domain of definition and values of the functions $\phi$ and $\theta$ is analyzed. For example, if the domains are $\mathbb{R} \times \mathbb{R}^{3}$ and $\mathbb{C}$, the maps act upon the arbitrary distant point relative to the origin, which implies that regularity conditions on the type $\vert \phi
(x^{0} , \mathbf{x} ) \vert \to 0$ and $\vert \theta
(x^{0} , \mathbf{x} ) \vert \to 0$ at $\vert \mathbf{x} \vert \to \infty$  must be imposed. Another consequence is that in the calculations, the limits $\vert \phi \vert \gg 1$ and  $\vert \theta \vert \gg 1$ are accessible. On the other hand, if the maps have compact subdomains such as 
$\mathbb{R} \times S^{3}$ and $S^2$ which is the case of the Hopf knots, the regularity conditions and the limits are different.

Another important property of the GEM knot fields discussed here is the existence of the constraints that must be imposed on the scalar maps due to the gravitational nature of the GEM fields, i.e., Equations (\refeq{GEM-gauge-cond}) above. While the Lorentz gauge is a natural relation among the potential vector components in electromagnetism and, therefore, leads to known properties of the knot fields, the pure GEM constraint is new and emerges in several places in GEM literature. These equations establish conditions that the knot maps must satisfy, which are new in the knot fields study. To exemplify, let us consider the second equation from 
(\refeq{GEM-gauge-cond}) with the second equation for the ansatz (\refeq{GEM-ansatze}) in the polar representation. It is easy to see that the following set of equations should be satisfied

\begin{align}
\left( \rho^{2} + 1 \right) \rho \partial_{0} \partial^{i} \alpha 
+
2 \partial_{0} \rho \partial^{i} \alpha
& = 0
\, ,
\label{GEM-constraint-phi}
\\
\left( \lambda^{2} + 1 \right) \lambda \partial_{0} \partial^{i} \beta 
+
2 \partial_{0} \lambda \partial^{i} \beta
& = 0
\, .
\label{GEM-constraint-theta}
\end{align}
Thus, we conclude that given one of the two real scalars that determines either the potential $\mathsf{A}$ or $\mathsf{C}$, one can determine the second scalar. To understand these equations' content, consider the variation in $x^{0}$ at a fixed point in space $x^{i}$. Then a simple calculation shows that (\refeq{GEM-constraint-phi}) and (\refeq{GEM-constraint-theta}) imposes the following constraint among the functions
$(\rho, \alpha )$ and $(\lambda, \beta)$

\begin{equation}
\delta_{ij} \partial^{i} \alpha \partial^{j} \alpha 
= 
\frac{3 \left(1+ \rho^{2}\right)}{\rho^{4}}
\, ,
\qquad
\delta_{ij} \partial^{i} \beta \partial^{j} \beta 
= 
\frac{3 \left(1+ \lambda^{2}\right)}{\lambda^{4}}
\, .
\label{GEM-constraints-solutions}
\end{equation}
The above equations place the spatial derivatives of the phases of the
maps $\phi$ and $\theta$ on spheres whose radii results from calculating the modulus of these functions at any instant of time. 

The study of the analogy between classical electromagnetism and the GEM equations in the linearized gravity is an exciting research field. The study of the GEM knot fields started here unfolds significant and noteworthy problems. One important aspect worth being studied further is the stability of the knots fields in the presence of matter and under perturbations. The knots fields' stability in the presence of matter and under perturbations is an open problem even in classical electromagnetism. Furthermore, it is essential to investigate other mathematical and physical properties of the general knots such as their topological structure, the energy-momentum tensor (along the line of \cite{Mashhoon:1998tt}), the types of fields limited by constraints, the generations of new solutions, the discussion of concrete examples and the analysis of the GEM knot fields in different limits, to mention just a few. We hope to report on some of these topics soon.

\section*{Acknowledgements}

I. V. V. would like to thank O. Lechtenfeld for correspondence at the beginning of this project.

\end{document}